\documentclass[prd,english]{revtex4}
\usepackage[T1]{fontenc}
\usepackage[latin1]{inputenc}
\usepackage{graphicx}
\usepackage{amsmath}
\usepackage{graphicx}
\usepackage{babel}

\begin{document}

\title{Anomalous production of top quarks at CLIC+LHC based $\gamma p$
colliders}

\author{Orhan \c{C}ak\i r}

\email{ocakir@science.ankara.edu.tr}

\homepage{http://science.ankara.edu.tr/-ocakir}

\address{Ankara University, Faculty of Sciences, Department of Physics, 06100,
Tandogan, Ankara, Turkey. }

\begin{abstract}
The single production of top quark due to flavor changing neutral
current (FCNC) interaction and its decay to $bW$ are studied at
CLIC+LHC based $\gamma p$ colliders. We consider both $tc\gamma $
and $tu\gamma $ anomalous couplings. The anomalous charm (up)
quark anomalous coupling parameter $\kappa _{\gamma }^{c}$
($\kappa _{\gamma }^{u}$) can be probed down to $9.5\times
10^{-3}$ ($8.0\times 10^{-3}$) at a $\gamma p$ collider with
$\sqrt{s_{ep}}=6.48$ TeV and $L_{\textrm{int}}=100$
$\textrm{fb}^{-1}$.
\end{abstract}


\maketitle

The flavor changing neutral current (FCNC) reactions are known to
be absent at tree level in the Standard Model (SM). However, they
can naturally appear at the one-loop level due to CKM mixing which
leads to the branching ratio BR($t\rightarrow qV$)$\sim
10^{-13}-10^{-10}$ \cite{00} where $q=c,u$ and $V=\gamma,g,Z$.
Extensions of the SM such as supersymmetry (SUSY) \cite{01},
exotic quarks (EXQ) \cite{02},\cite{02_1} and two-Higgs doublet
models (2HDM) \cite{03} could lead to an enhancements of such
transitions. In these models the top quark is predicted to have
large FCNC couplings \cite{1,2}. The approximate orders of the
branching ratios for FCNC top quark decays predicted within these
models {[}see \cite{3} and references therein{]} are given in
Table \ref{table1}.

Due to the large mass close to the electroweak symmetry breaking scale
and having poorly measured couplings, the top quark is a good candidate
for probing new physics beyond the SM. This motivates the study of
single top production by FCNC couplings at future colliders. The production
of top quarks via FCNC interactions was extensively studied at hadron
colliders \cite{4,5,6,7,8,9,10}, at $e^{+}e^{-}$ colliders \cite{10,11,12,13},
and at lepton-hadron colliders \cite{13,14}.

Additional option of linear $e^+e^-$ colliders would be an ep
collider when linear collider is constructed on the same base as
the proton ring. The compact linear collider (CLIC) \cite{14_1} at
CERN can be converted into CLIC+LHC $ep$ collider. Photon
colliders \cite{14_2} are based on the Compton scattering of laser
light on high energy electrons in linear colliders. The scattered
photons have energies close to the energy of the initial electron
beams. Then, it is possible to construct TeV scale $\gamma p$
collider on the CLIC+LHC base. The main parameters of the CLIC+LHC
based $\gamma p$ colliders are given in Table \ref{table2}.

In this paper, single production of top quarks via anomalous
$tc\gamma $ and $tu\gamma $ couplings at future CLIC+LHC based
$\gamma p$ colliders has been studied.

If the FCNC couplings of the top quark exist, they will affect top
quark production and decay at high energies. Therefore, any possible
deviations from SM predictions will be an indication of new physics.
Top quark FCNC couplings can be parametrized in a model independent
way by an effective Lagrangian

\begin{eqnarray}
L & = & \sum _{q=u,c}ig_{e}Q_{q}\frac{\kappa _{\gamma }^{q}}{\Lambda }\overline{t}\sigma _{\mu \nu }k^{\nu }qA^{\mu } \nonumber \\
&  & +\sum _{q=u,c}\frac{g_{e}}{2\sin \theta _{W}\cos \theta _{W}}\overline{t}\left[i\frac{\kappa _{Z}^{q}}{\Lambda }\sigma _{\mu \nu }k^{\nu }-\gamma _{\mu }(v_{Z}^{q}-a_{Z}^{q}\gamma _{5})\right]qZ^{\mu } \nonumber \\
&  & +\sum _{q=u,c}ig_{s}\frac{\kappa _{g}^{q}}{\Lambda
}\overline{t}\sigma _{\mu \nu }k^{\nu }\frac{\lambda
^{a}}{2}qG_{a}^{\mu }+\textrm{H}.\textrm{c}.
\end{eqnarray}
where $k^{\nu }$ is the momentum of the neutral gauge boson,
$\sigma _{\mu \nu }=i(\gamma _{\mu }\gamma _{\nu }-\gamma _{\nu
}\gamma _{\mu })/2$ and $\Lambda $ is the new physics scale.
$A^{\mu }$, $Z^{\mu }$ and $G^{\mu }$ are the photon, Z- boson and
gluon fields, respectively. The anomalous couplings $\kappa
_{\gamma },\textrm{ }\kappa _{Z}\textrm{ ve }\kappa _{g}$ define
the strength of the $tq\gamma $, $tqZ$ and $tqg$ vertices,
respectively. The terms including $v_{Z}$ and $a_{Z}$ are the
anomalous non-diagonal $Z$ couplings which are zero in the SM.
$g_{e}$ and $g_{s}$ are the electromagnetic and strong coupling
constants, respectively. $Q_{q}$ is the quark charge and $\theta
_{W}$ is the Weinberg angle.

Using the effective Lagrangian (1) it is straightforward to obtain
the FCNC decay widths of top quark:

\begin{equation}
\Gamma (t\rightarrow qg)=\left(\frac{\kappa _{g}^{q}}{\Lambda
}\right)^{2}\frac{2}{3}\alpha _{s}m_{t}^{3}
\end{equation}

\begin{equation}
\Gamma (t\rightarrow q\gamma )=\left(\frac{\kappa _{\gamma
}^{q}}{\Lambda }\right)^{2}\frac{2}{9}\alpha
m_{t}^{3}
\end{equation}

\begin{equation}
 \Gamma (t\rightarrow qZ)=\frac{\alpha m_{t}}{4\sin ^{2}2\theta
_{W}}\left(1-\frac{m_{Z}^{2}}{m_{t}^{2}}\right)^{2}\left[\left(\frac{\kappa
_{Z}^{q}}{\Lambda
}\right)^{2}m_{t}^{2}(2+\frac{m_{Z}^{2}}{m_{t}^{2}})-6v_{Z}^{q}\frac{\kappa
_{Z}^{q}}{\Lambda
}m_{t}+(v_{Z}^{q2}+a_{Z}^{q2})(2+\frac{m_{t}^{2}}{m_{Z}^{2}})\right]
\end{equation}
where $m_{Z}$ and $m_{t}$ are the masses of Z boson and top quark,
respectively.

The CDF collaboration performed a search for FCNC in the top quark
decays $t\rightarrow u(c)\gamma $ and $t\rightarrow u(c)Z$ in $p\overline{p}$
collisions at a centre of mass energy of $1.8$ TeV. They obtained
upper limits at 95\% confidence level (CL) on the branching ratios
\cite{15}:

\[
BR(t\rightarrow u\gamma )+BR(t\rightarrow c\gamma )<0.032,
\]
\[
BR(t\rightarrow uZ)+BR(t\rightarrow cZ)<0.33.
\]

A slightly better limit from the DELPHI experiments for the
process $e^{+}e^{-}\rightarrow t\overline{q}$ is obtained on the
top quark FCNC branching fraction $BR(t\rightarrow
uZ)+BR(t\rightarrow cZ)<0.18$ \cite{16} assuming other FCNC
couplings to be zero. Note that the LEP2 gives a better limit for
$\kappa _{Z}^{q}<0.52$ and
$\sqrt{|v_{Z}^{q}|^{2}+|a_{Z}^{q}|^{2}}<0.32$ than that given by
the CDF. From the low energy experiments a more stringent upper
bound on the $tqZ$ couplings $\sqrt{|v_{Z}|^{2}+|a_{Z}|^{2}}<0.15$
can be found in \cite{3}.

The H1 Collaboration searched for the production processes by
considering both of the anomalous vertices $tu\gamma $ and
$tc\gamma $ from the fact that HERA has much higher sensitivity to
$\kappa _{\gamma }^{u}$ than to $\kappa _{\gamma }^{c}$, due to
more favorable parton density \cite{17}. The current limits from
HERA are $\kappa _{\gamma }^{q}<0.19$ (ZEUS) and $\kappa _{\gamma
}^{q}<0.305$ (H1) \cite{18}.

In this study we define the branching ratio for FCNC decay of top
quark as
\begin{equation} BR(t\rightarrow qV)=\frac{\Gamma
(t\rightarrow qV)}{\sum \Gamma (t\rightarrow qV)}.
\end{equation}

\begin{figure}[htbp]
\begin{center}\includegraphics{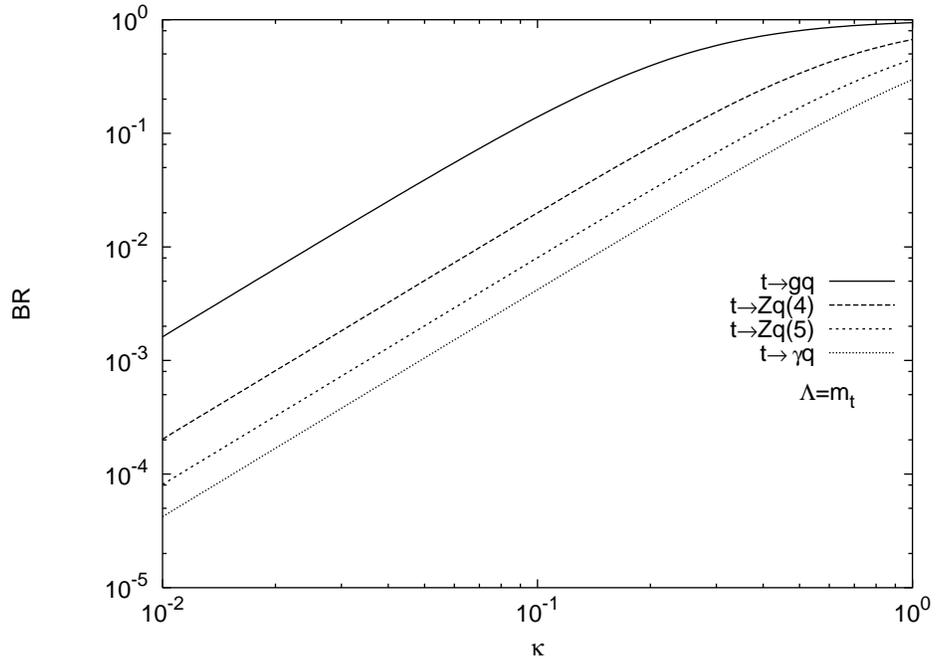}\end{center}
\caption{Branching ratios depending on the anomalous coupling
$\kappa .$ Only the relevant FCNC coupling is allowed to deviate
from its SM value at a time as the others are set to zero. The
branchings $t\rightarrow qZ$ are considered in two cases as
explained in the text.\label{fig1}}
\end{figure}

By convention, we set $\Lambda =m_{t}=175$ GeV and $\alpha
=1/128$, $\alpha _{s}=\alpha _{s}(m_{t}^{2})$ in our calculations.
The branching ratios BR of $t\rightarrow qg$, $t\rightarrow qZ$
and $t\rightarrow q\gamma $ are shown in Fig. \ref{fig1}. Using
the effective lagrangian (1) with the scale $\Lambda =m_{t}=175$
GeV, and standard decay width $\Gamma (t\rightarrow bW)\simeq 1.4$
GeV, these branchings can be translated into the bounds on the top
quark anomalous couplings $\kappa _{\gamma }^{q}<0.28$, $\kappa
_{Z}^{q}<0.77$ and $\sqrt{|v_{Z}^{q}|^{2}+|a_{Z}^{q}|^{2}}<0.48$
as seen in Fig. \ref{fig1}.

The branchings for $t\rightarrow qZ$ are denoted by the
contribution from a 5-dimensional FCNC coupling $\kappa $ or the
4-dimensional FCNC couplings
$\sqrt{|v_{Z}^{q}|^{2}+|a_{Z}^{q}|^{2}}=\kappa $ for an
illustration. Here, we assume that only the relevant FCNC coupling
is allowed to deviate from its SM value at a time as the others
are set to zero.

Since we allow $tc\gamma$ and $tu\gamma$ anomalous interaction in
the direct top quark production we should also include this term
in the decay of top quark, which is proportional to $(\kappa
_{\gamma }^{c})^{2}+(\kappa _{\gamma }^{u})^{2}$ and contributes
to the top quark decay width. One can see the effect of additional
channel for top quark decay, which decreases the $t\rightarrow bW$
branching ratio and causes a noticeable deviation from the
quadratic behavior. Assuming only the presence of $tc\gamma$ and
$tu\gamma$ couplings top quark decay width changes according to
\begin{equation}
\Gamma_t=\Gamma_{t\to
bW}\left[1+{[(\kappa_\gamma^{c})^2+(\kappa_\gamma^{u})^2]\over
\Lambda^2}{32\over 9}{m_W^2\over
(1-m_W^2/m_t^2)^2(1+2m_W^2/m_t^2)}\right].
\end{equation}

While the $t\to qV$ ($q=u,c$ and $V=\gamma,Z,g$) decays will occur
in the presence of the anomalous couplings given in Eq. (1), they
are smaller than the $t\to bW$ decay and they will have negligible
branching ratios for $\kappa_V<0.2$ at $\Lambda =m_{t}$. Given the
existing upper bound of the anomalous coupling mentioned earlier
\cite{4}, $t\rightarrow bW$ will be the dominant decay mode of the
top quark. Since the leptonic decay channel of W boson has a clear
signature, here we consider only the $t\rightarrow bW^+\rightarrow
l^+\nu b$ decay for our signal.

The total cross section for direct top quark production is given
by

\begin{equation}
\sigma =\int_{\tau_{\textrm{min}}}^{0.83}\int
_{\tau/0.83}^{1}{dx\over x}\: f_{q}(x,Q^{2})f_{\gamma}({\tau\over
x})\int _{\hat{t}_{-}}^{\hat{t}_{+}}d\hat{t}\frac{d\hat{\sigma
}}{d\hat{t}}
\end{equation}
where $\hat{t}_{\textrm{-}}=(m_{t}^{2}-\hat{s})$, $\hat{t}_{+}=0$
and $\tau_{\textrm{min}}=m_{t}^{2}/s$. The $f_\gamma(y)$ is the
spectrum of photons scattered backward from the interaction of
laser photon with the high energy electrons \cite{ginzburg}. The
differential cross section for the subprocess $\gamma q\rightarrow
t\rightarrow bW^+$ is given by
\begin{eqnarray}
\frac{d\hat{\sigma }}{d\hat{t}} & = & \left(\frac{\kappa _{\gamma
}^{q}}{\Lambda }\right)^{2} \frac{\pi \alpha
^{2}|V_{tb}|^{2}}{9m_{w}^{2}\textrm{sin}^{2}\theta
_{w}\hat{s}[(\hat s-m_t^2)^2+m_t^2\Gamma_t^2]}
\left[\hat{s}^{3}+\hat{s}^{2}\hat{t}-(m_{w}^{2}+2m_{b}^{2})
\hat{s}^{2}-(2m_{w}^{2}+m_{t}^{2}+m_{b}^{2})\hat{s}\hat{t}\right.\nonumber \\
 &  & \left.+(m_{b}^{4}+2m_{b}^{2}m_{w}^{2}+2m_{t}^{2}m_{w}^{2})
 \hat{s}+(m_{b}^{2}m_{t}^{2}+2m_{t}^{2}m_{w}^{2})
 \hat{t}-m_{b}^{2}m_{t}^{2}m_{w}^{2}-2m_{t}^{2}m_{w}^{4}\right]
\end{eqnarray}

\begin{figure}[htbp]
\begin{center}\includegraphics[  scale=0.6]{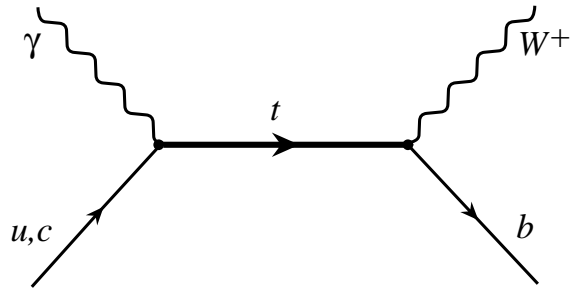}\end{center}
\caption{Feynman graph for anomalous top quark production in
$\gamma p$ collision.\label{fig2}}
\end{figure}

In this study, the anomalous interaction vertices are implemented
into the CALCHEP \cite{18} package with the parton distribution
function library CTEQ5M \cite{19} for $f_{q}(x,Q^{2})$ at
$Q^{2}=m_{t}^{2}$. The FCNC $tq\gamma $ couplings can be probed
directly at $\gamma p$ colliders through the top quark production
subprocesses $\gamma u\rightarrow t$ and $\gamma c\rightarrow t$.
The diagram for the FCNC top quark production mechanism and its
subsequent decay $t\rightarrow bW^+\rightarrow l^+\nu b$ are shown
in Fig. \ref{fig2}.

The cross sections for anomalous top quark production at three
options of $ep$ colliders with the center of mass energies
$\sqrt{s_{ep}}=2.64$, $3.74$ and $6.48$ TeV are given in Tables
\ref{table3}, \ref{table3} and \ref{table5}, respectively. From
these tables it is easy to see that the contribution from
$tu\gamma $ anomalous coupling is larger than that from $tc\gamma
$ coupling.

\begin{figure}[htbp]
\begin{center}\includegraphics{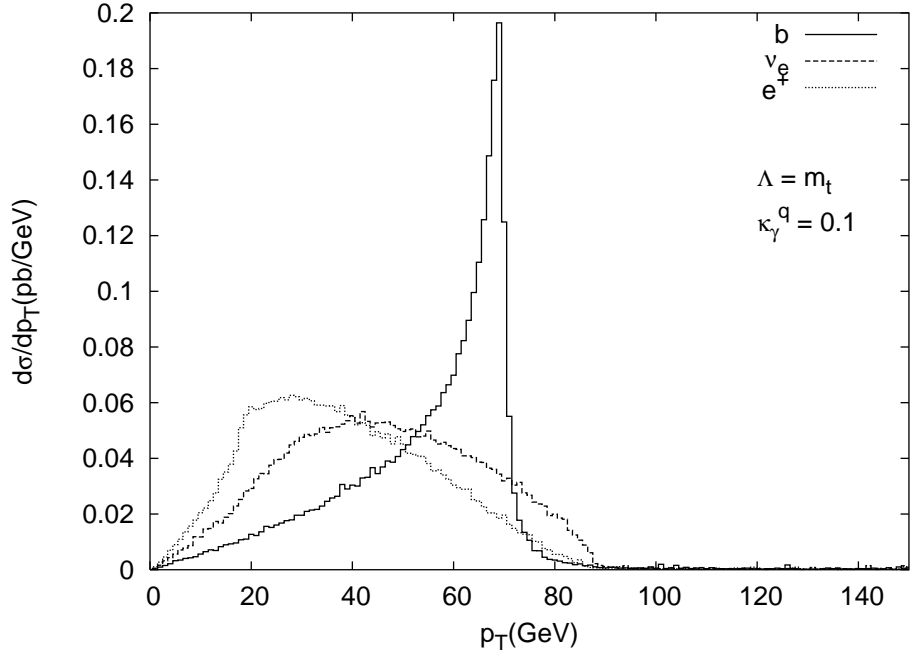}\end{center}
\caption{The transverse momentum distribution of $b$-quark, lepton
($e^{+}$ or $\mu ^{+}$), and the missing $p_{T}$ from the FCNC top
production at a $\gamma p$ collider based on CLIC+LHC with
$\sqrt{s_{\textrm{ep}}}=3.74$ TeV for the scale $\Lambda =m_{t}$
and the anomalous coupling $\kappa _{\gamma}^q=0.1$. \label{fig3}}
\end{figure}

\begin{figure}[htbp]
\begin{center}\includegraphics{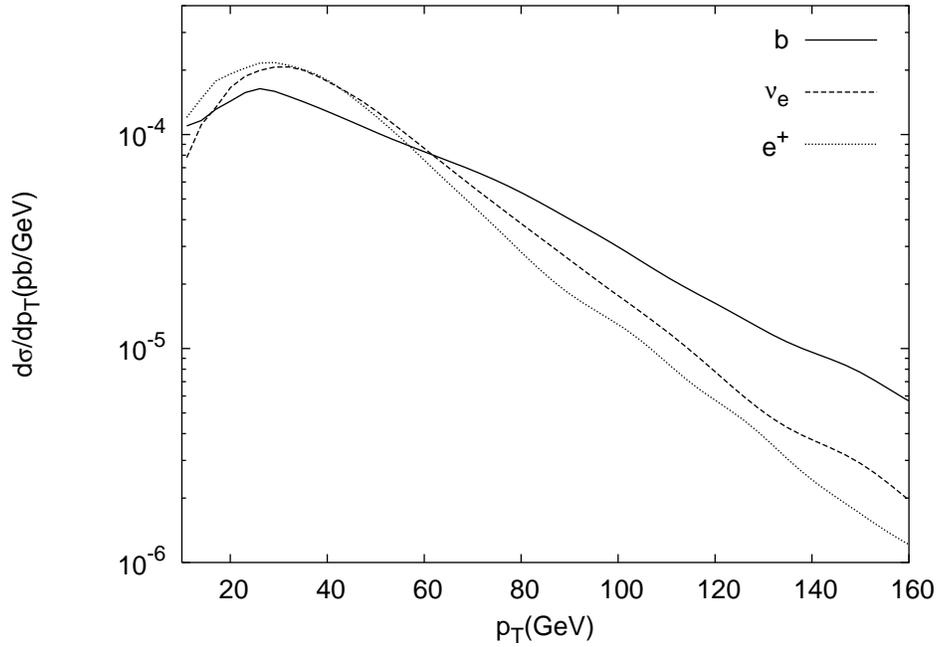}\end{center}
\caption{The transverse momentum distribution of $b$-quark, lepton
($e^{+}$ or $\mu ^{+}$), and the missing $p_{T}$ for the process
$ep\rightarrow e^{+}\nu bX$ at a $\gamma p$ collider with
$\sqrt{s_{\textrm{ep}}}=3.74$ TeV. \label{fig4}}
\end{figure}

We search for the signal in detector through the presence of one
$b$-tagged jet, one isolated lepton and missing transverse
momentum. The transverse momentum $p_{T}$ distribution of final
state particles are given in Fig. \ref{fig3}. The $b$ quark $p_T$
distribution has a peak around the half mass of top quark. A small
shift backwards due to the mass of final state particles is seen.
The $p_T$ spectrum of the electron (or positron) mainly
distributed at the half mass of the $W$ boson. In Fig. \ref{fig4}
we present the $p_T$ spectrum of final state particles for the
relevant background.

We apply initial kinematic cuts $p_{T}^{e,\nu ,b}>10$ GeV for the
experimental observation of the signal. These cuts reduce the
signal cross section by 5\% and background by 45\%. More stringent
cuts $p_{T}^{e,\nu }>20$ GeV and $p_{T}^{b}>50$ GeV lead to a
reduction on the signal cross section about 50\% and background
about 85\%. The rapidity cuts are more effective such that $|\eta
^{e,\nu ,b}|<2.5$ reduces the signal cross section by a factor
about $80\%$ and background by $90\%$.

\begin{figure}[htbp]
\begin{center}\includegraphics{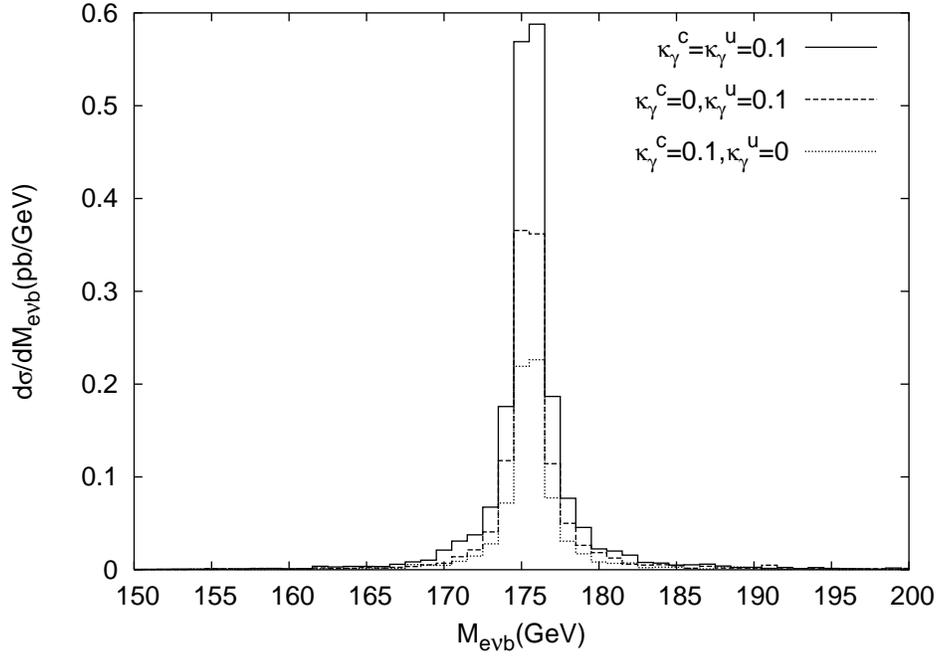}\end{center}
\caption{$M_{e\nu b}$ distributions for the cut $p_{T}^{e,\nu
,b}>20$ GeV at the $\gamma p$ collider with
$\sqrt{s_{ep}}=3.74$TeV. \label{fig5}}
\end{figure}

In order to enhance the signal to the background, we want to make
cuts on the invariant mass of final particles which should be sharply
peaked at $m_{t}$ for the signal. To determine the invariant mass
$M_{l\nu b}$, one should reconstruct $p_{t}=p_{l}+p_{\nu }+p_{b}$.
The neutrino is not observed but its transverse momentum can be deduced
from the missing transverse momentum. The longitudinal component of
the neutrino momentum is determined by the W-mass constraint
$m_{W}=m_{l\nu }$=$\sqrt{(p_{l}+p_{\nu })^{2}}$,
and is given by

\begin{equation}
p_{L}^{\nu }=\frac{\chi p_{L}^{l}\pm
\sqrt{p_{l}^{2}(\chi^{2}-p_{Tl}^{2}p_{T\nu
}^{2})}}{p_{Tl}^{2}},\qquad \textrm{ }\chi
=\frac{m_{W}^{2}}{2}+p_{T}^{l}\cdot p_{T}^{\nu }
\end{equation}
and $p_{L}$ and $p_{T}$ refer to the longitudinal and transverse
momenta, respectively. We chose the solution which would best
reconstruct the mass of the top quark. The invariant mass
distributions of $l\nu b$ system for the three options of
anomalous interaction vertex parameters $\kappa_\gamma^c$ and
$\kappa_\gamma^u$ are shown in Fig. \ref{fig5} where we apply the
cuts $p_{T}^{l,\nu ,b}>20$ GeV.

We can determine the minimum value of the anomalous couplings
$\kappa_\gamma^c$ or $\kappa_\gamma^u$ by using the signal and
background events at $\gamma p$ colliders. Assuming the Poisson
statistics the number of signal events required for discovery of a
signal at the $95\%$ confidence level is

\begin{equation}
{S\over \sqrt{S+B}}\geq 3
\end{equation}
where $S$ and $B$ are defined according to the number of events
calculation $(S,B)=\sigma\times BR_w\times L_{int}\times \epsilon
$, where $\epsilon$ is the overall detection efficiency of $1\%$
for this channel. $L_{int}$ is the integrated luminosity for one
working year, and $BR_w$ is the branching ratio for leptonic decay
of the $W$ boson.

\begin{figure}[htbp]
\begin{center}\includegraphics{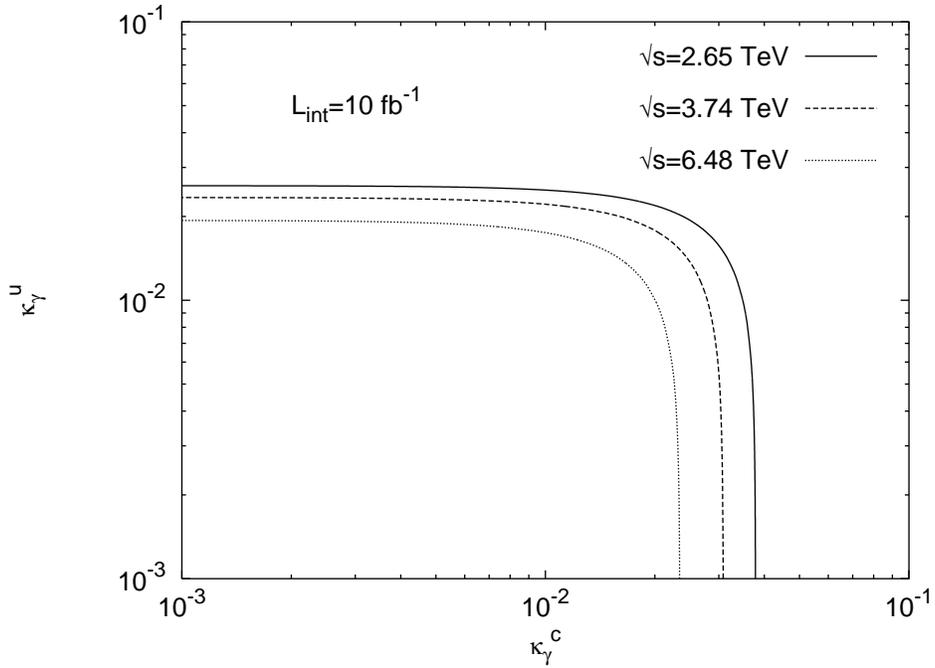}\end{center}
\caption{Discovery limits for $\kappa ^{c}$ and $\kappa ^{u}$ for
each of the colliders considered at $L=10\textrm{
fb}^{\textrm{-1}}$, for $\Lambda =m_{t}$. \label{fig6}}
\end{figure}

\begin{figure}[htbp]
\begin{center}\includegraphics{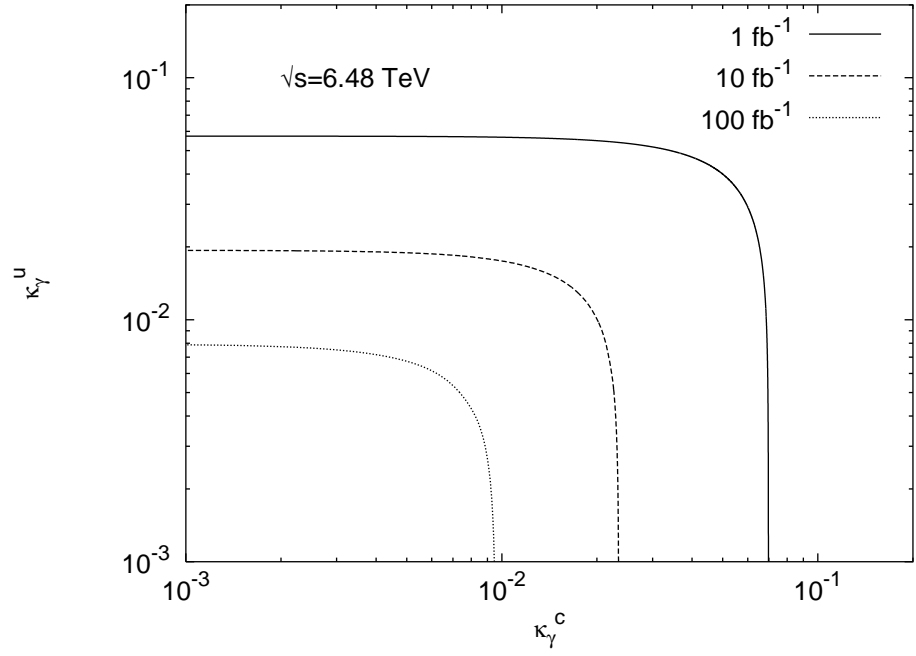}\end{center}
\caption{Discovery limits for $\kappa ^{c}$ and $\kappa ^{u}$ at
CLIC+LHC based $\gamma p$ colliders at $\sqrt{s}=6.48$ TeV for
three different luminosity and $\Lambda =m_{t}$. \label{fig7}}
\end{figure}

Since the charm and up quarks are in the initial state, their
contributions to direct top quark production can not be
distinguished. The plots of the discovery limit when both
$\kappa_\gamma^c$ and $\kappa_\gamma^u$ are assumed to be nonzero
are shown in Figs. \ref{fig6} and \ref{fig7}.

We find that cross section is more sensitive to $tu\gamma $ than
$tc\gamma $ due to the more favorable $u$ quark density in proton.
From Fig. \ref{fig6} one can see that larger center of mass energy
improves the sensitivity. Assuming the integrated luminosity of
$100$ fb$^{-1}$ one can probe the anomalous $tc\gamma$
($tu\gamma$) couplings down to the value of 0.0095 (0.008) when
the scale of the interaction is set to the top quark mass at a
CLIC+LHC based $\gamma p$ collider with $\sqrt{s_{\gamma
p}^{max}}\simeq 5.9$ TeV. For other choices of $\Lambda$ the
results can be rescaled by $(m_t/\Lambda)^2$. These limits
correspond to the branching ratio $BR\to q\gamma\approx 5\times
10^{-5}$. In the SM extension with exotic quarks
\cite{02},\cite{02_1} and supersymmetric models with R-parity
violation \cite{01}, this branching ratio can be as low as
$O(10^{-5})$. If these models are the only source for the
anomalous $tq\gamma$ couplings, our calculations therefore
indicate that future improvements at the CLIC+LHC based $\gamma p$
colliders will be needed to make this a detectible signal unless
LO and NLO corrections further enhance the contributions.

In principle, there can be the overlapping between the direct top
quark production and the photon or gluon splitting diagrams where
photon (gluon) splits into a $q\bar{q}$ pair, and $u$ and/or $c$
combines with the gluon (photon) to produce a top quark. Care must
be taken with these processes to avoid the double counting. In the
direct production due to anomalous $tc\gamma $ and/or $tu\gamma $
couplings, initial state particles are assumed to be massless, and
thus we may ignore the effects of the double counting problem.

In conclusion, the anomalous couplings can be large in a specific
model. The present experimental limits are relatively weak and
these couplings can exist in tree level anomalous processes and
can be measured with a better precision at high energy
lepton-hadron colliders. The FCNC interaction of top quark would
be a signal for the existence of new physics beyond the SM.

\begin{table}[htbp]
\caption{Branching ratios for FCNC top quark decays as predicted
within the SM and the SM extensions.\label{table1} }
\begin{center}\begin{tabular}{cccc}
\hline
&
 $BR(t\rightarrow qg)$&
 $BR(t\rightarrow q\gamma )$&
 $BR(t\rightarrow qZ)$\\
\hline
SM&
 10$^{-10}$&
 10$^{-12}$&
 10$^{-13}$\\
 2HDM&
 10$^{-5}$&
 10$^{-7}$&
 10$^{-6}$\\
 SUSY&
 10$^{-5}$&
 10$^{-6}$&
 10$^{-6}$\\
 EXQ&
 10$^{-3}$&
 10$^{-5}$&
 10$^{-2}$ \\
\hline
\end{tabular}\end{center}
\end{table}

\begin{table}[htbp]
\caption{The main parameters of the options of CLIC+LHC based
$\gamma p$ colliders. \label{table2} }
\begin{center}\begin{tabular}{|c|c|c|c|c|c|}
\hline
CLIC+LHC&
 $E_{e}(\textrm{TeV})$&
 $E_{p}(\textrm{TeV})$&
 $\sqrt{s_{ep}}$(TeV)&
 $\sqrt{s_{\gamma p}^{max}}$(TeV)&
 $L_{ep}\simeq L_{\gamma p}(\times 10^{32}\textrm{cm}^{-2}\textrm{s}^{-1})$\\
\hline
Option 1&
 0.25&
 7&
 2.64&
 2.41&
1-100 \\
\hline
Option 2&
 0.50&
 7&
 3.74&
 3.41&
1-100\\
\hline
Option 3&
 1.50&
 7&
 6.48&
 5.90&
 1-100\\
\hline
\end{tabular}\end{center}
\end{table}

\begin{table*}[htbp]
\caption{The cross section for the process $\gamma p\rightarrow
bWX$ at a CLIC+LHC based $\gamma p$ collider with
$\sqrt{s_{ep}}=2.64$ TeV.\label{table3}}
\begin{tabular}{|c|c|c|c|c|c|c|c|c|}
\hline
\multicolumn{2}{|c|}{$\Lambda =m_{t}$}&
\multicolumn{7}{c|}{$\kappa _{\gamma }^{u}$ }\\
\cline{3-9}
\multicolumn{2}{|c|}{ $\sigma $(pb)}&
 0&
 0.01&
 0.05&
 0.1&
 0.2&
 0.3&
 0.4\\
\hline
\multicolumn{1}{|c|}{}&
 0&
 $4.95\times 10^{-2}$&
 $1.92\times 10^{-1}$&
 $3.61\times 10^{0}$&
 $1.43\times 10^{1}$&
 $5.69\times 10^{1}$&
 $1.28\times 10^{2}$&
 $2.82\times 10^{2}$\\
\multicolumn{1}{|c|}{}&
 0.01&
 $1.16\times 10^{-1}$&
 $3.08\times 10^{-1}$&
 $3.73\times 10^{0}$&
 $1.44\times 10^{1}$&
 $5.70\times 10^{1}$&
 $1.28\times 10^{2}$&
 $2.82\times 10^{2}$\\
\multicolumn{1}{|c|}{}&
 0.05&
 $1.70\times 10^{0}$&
 $2.08\times 10^{0}$&
 $5.50\times 10^{0}$&
 $1.62\times 10^{1}$&
 $5.88\times 10^{1}$&
 $1.29\times 10^{2}$&
 $2.84\times 10^{2}$\\
\multicolumn{1}{|c|}{$\kappa _{\gamma }^{c}$}&
 0.1&
 $6.67\times 10^{0}$&
 $6.86\times 10^{0}$&
 $1.03\times 10^{1}$&
 $2.09\times 10^{1}$&
 $6.36\times 10^{1}$&
 $1.35\times 10^{2}$&
 $2.88\times 10^{2}$\\
\multicolumn{1}{|c|}{}&
 0.2&
 $2.65\times 10^{1}$&
 $2.67\times 10^{1}$&
 $6.68\times 10^{1}$&
 $8.43\times 10^{1}$&
 $1.54\times 10^{2}$&
 $2.69\times 10^{2}$&
 $4.34\times 10^{2}$\\
\multicolumn{1}{|c|}{}&
 0.3&
 $5.95\times 10^{1}$&
 $1.37\times 10^{2}$&
 $1.43\times 10^{2}$&
 $1.60\times 10^{2}$&
 $2.30\times 10^{2}$&
 $3.46\times 10^{2}$&
 $5.10\times 10^{2}$\\
\multicolumn{1}{|c|}{}&
 0.4&
 $1.06\times 10^{2}$&
 $2.43\times 10^{2}$&
 $2.49\times 10^{2}$&
 $2.66\times 10^{2}$&
 $3.36\times 10^{2}$&
 $4.52\times 10^{2}$&
 $6.16\times 10^{2}$ \\
\hline
\end{tabular}
\end{table*}

\begin{table*}[htbp]
\caption{The cross section for the process $\gamma p\rightarrow
bWX$ at a CLIC+LHC based $\gamma p$ collider with
$\sqrt{s_{ep}}=3.74$ TeV. \label{table4} }
\begin{tabular}{|c|c|c|c|c|c|c|c|c|}
\hline
\multicolumn{2}{|c|}{$\Lambda =m_{t}$}&
\multicolumn{7}{c|}{$\kappa _{\gamma }^{u}$ }\\
\cline{3-9}
\multicolumn{2}{|c|}{ $\sigma $(pb)}&
 0&
 0.01&
 0.05&
 0.1&
 0.2&
 0.3&
 0.4\\
\hline
\multicolumn{1}{|c|}{}&
 0&
 $8.12\times 10^{-2}$&
 $2.59\times 10^{-1}$&
 $4.52\times 10^{0}$&
 $1.78\times 10^{1}$&
 $7.11\times 10^{1}$&
 $1.59\times 10^{2}$&
 $2.84\times 10^{2}$\\
\multicolumn{1}{|c|}{}&
 0.01&
 $1.84\times 10^{-1}$&
 $4.43\times 10^{-1}$&
 $4.70\times 10^{0}$&
 $1.79\times 10^{1}$&
 $7.13\times 10^{1}$&
 $1.59\times 10^{2}$&
 $2.84\times 10^{2}$\\
\multicolumn{1}{|c|}{}&
 0.05&
 $2.63\times 10^{0}$&
 $2.89\times 10^{0}$&
 $7.15\times 10^{0}$&
 $2.04\times 10^{1}$&
 $7.37\times 10^{1}$&
 $1.62\times 10^{2}$&
 $2.87\times 10^{2}$\\
\multicolumn{1}{|c|}{$\kappa _{\gamma }^{c}$}&
 0.1&
 $1.03\times 10^{1}$&
 $1.05\times 10^{1}$&
 $1.48\times 10^{1}$&
 $2.81\times 10^{1}$&
 $8.14\times 10^{1}$&
 $1.69\times 10^{2}$&
 $2.94\times 10^{2}$\\
\multicolumn{1}{|c|}{}&
 0.2&
 $4.10\times 10^{1}$&
 $4.12\times 10^{1}$&
 $4.55\times 10^{1}$&
 $5.88\times 10^{1}$&
 $1.12\times 10^{2}$&
 $2.00\times 10^{2}$&
 $3.25\times 10^{2}$\\
\multicolumn{1}{|c|}{}&
 0.3&
 $9.22\times 10^{1}$&
 $9.24\times 10^{1}$&
 $9.67\times 10^{1}$&
 $1.10\times 10^{2}$&
 $1.63\times 10^{2}$&
 2.51$\times 10^{2}$&
 $3.76\times 10^{2}$\\
\multicolumn{1}{|c|}{}&
 0.4&
 $1.64\times 10^{2}$&
 $1.64\times 10^{2}$&
 $1.68\times 10^{2}$&
 $1.82\times 10^{2}$&
 $2.35\times 10^{2}$&
 $3.23\times 10^{2}$&
 $4.48\times 10^{2}$ \\
\hline
\end{tabular}
\end{table*}

\begin{table*}[htbp]
\caption{The cross section for the process $\gamma p\rightarrow
bWX$ at a CLIC+LHC based $\gamma p$ collider with
$\sqrt{s_{ep}}=6.48$ TeV.\label{table5} }
\begin{tabular}{|c|c|c|c|c|c|c|c|c|}
\hline
\multicolumn{2}{|c|}{$\Lambda =m_{t}$}&
\multicolumn{7}{c|}{$\kappa _{\gamma }^{u}$ }\\
\cline{3-9}
\multicolumn{2}{|c|}{ $\sigma $(pb)}&
 0&
 0.01&
 0.05&
 0.1&
 0.2&
 0.3&
 0.4\\
\hline
\multicolumn{1}{|c|}{}&
 0&
 $2.45\times 10^{-1}$&
 $6.33\times 10^{-1}$&
 $9.95\times 10^{0}$&
 $3.91\times 10^{1}$&
 $1.55\times 10^{2}$&
 $3.50\times 10^{2}$&
 $6.21\times 10^{2}$\\
\multicolumn{1}{|c|}{}&
 0.01&
 $5.16\times 10^{-1}$&
 $1.15\times 10^{0}$&
 $1.05\times 10^{1}$&
 $3.96\times 10^{1}$&
 $1.55\times 10^{2}$&
 $3.50\times 10^{2}$&
 $6.21\times 10^{2}$\\
\multicolumn{1}{|c|}{}&
 0.05&
 $6.95\times 10^{0}$&
 $7.58\times 10^{0}$&
 $1.69\times 10^{1}$&
 $4.60\times 10^{1}$&
 $1.62\times 10^{2}$&
 $3.57\times 10^{2}$&
 $6.28\times 10^{2}$\\
\multicolumn{1}{|c|}{$\kappa _{\gamma }^{c}$}&
 0.1&
 $2.71\times 10^{1}$&
 $2.77\times 10^{1}$&
 $3.70\times 10^{1}$&
 $6.62\times 10^{1}$&
 $1.82\times 10^{2}$&
 $3.77\times 10^{2}$&
 $6.48\times 10^{2}$\\
\multicolumn{1}{|c|}{}&
 0.2&
 $1.07\times 10^{2}$&
 $1.08\times 10^{2}$&
 $1.17\times 10^{2}$&
 $1.46\times 10^{2}$&
 $2.62\times 10^{2}$&
 $4.57\times 10^{2}$&
 $7.28\times 10^{2}$\\
\multicolumn{1}{|c|}{}&
 0.3&
 $2.42\times 10^{2}$&
 $2.43\times 10^{2}$&
 $2.52\times 10^{2}$&
 $2.81\times 10^{2}$&
 $3.97\times 10^{2}$&
 $5.92\times 10^{2}$&
 $8.63\times 10^{2}$\\
\multicolumn{1}{|c|}{}&
 0.4&
 $4.29\times 10^{2}$&
 $4.29\times 10^{2}$&
 $4.39\times 10^{2}$&
 $4.68\times 10^{2}$&
 $5.84\times 10^{2}$&
 $7.79\times 10^{2}$&
 $1.05\times 10^{3}$ \\
\hline
\end{tabular}
\end{table*}

\end{document}